\documentclass[10pt]{iopart}
\eqnobysec
\usepackage{iopams}
\usepackage{graphicx}
\usepackage{siunitx}
\usepackage{braket}
\usepackage{cases}
\begin{document}

\title[Large-scale quantum socket]{Mitigating coherent leakage of superconducting qubits in a large-scale quantum socket}

\author{T G~McConkey$^{1,2}$, J H~B\'{e}janin$^{1,3}$, C T~Earnest$^{1,3}$, C R H~McRae$^{1,3}$, Z~Pagel$^{1,3}$\footnote{Present address: Department of Physics, University of California, Berkeley, 366 LeConte Hall, MC 7300 Berkeley, California 94720-7300, USA}, J R~Rinehart$^{1,3}$ and M~Mariantoni$^{1,3}$}

\address{$^{1}$ Institute for Quantum Computing, University of Waterloo, 200 University Avenue West, Waterloo, Ontario N2L 3G1, Canada}
\address{$^{2}$ Department of Electrical and Computer Engineering, University of Waterloo, 200 University Avenue West, Waterloo, Ontario N2L 3G1, Canada}
\address{$^{3}$ Department of Physics and Astronomy, University of Waterloo, 200 University Avenue West, Waterloo, Ontario N2L 3G1, Canada}

\ead{matteo.mariantoni@uwaterloo.ca}
\vspace{10pt}

\begin{indented}
\item[] October 12th, 2017
\end{indented}

\begin{abstract}
A practical quantum computer requires quantum bit~(qubit) operations with low error rates in extensible architectures. We study a packaging method that makes it possible to address hundreds of superconducting qubits by means of three-dimensional wires: \textit{The large-scale quantum socket}. A qubit chip is housed in a superconducting box, where both box and chip dimensions lead to unwanted modes that can interfere with qubit operations. We theoretically analyze these interference effects in the context of qubit coherent leakage. We propose two methods to mitigate the resulting errors by detuning the resonance frequency of the modes from the qubit frequency. We perform detailed electromagnetic field simulations indicating that the resonance frequency of the modes increases with the number of installed three-dimensional wires and can be engineered to be significantly higher than the highest qubit frequency. Finally, we show preliminary experimental results towards the implementation of a large-scale quantum socket.
\end{abstract}


\vspace{2pc}

\maketitle

\ioptwocol

\section{Introduction}
	\label{sec:Introduction}

The implementation of a practical quantum computer will make it possible to run certain algorithms much more efficiently than any classical computer~\cite{Nielsen:2000}. Search and optimization as well as cryptography algorithms, such as Shor's algorithm, will have profound implications on society at large~\cite{Montanaro:2016}. Additionally, digital quantum simulation algorithms will have an important impact on scientific research~\cite{Wecker:2015, Reiher:2017}.

Many implementations of the quantum computer and its primary component, the quantum bit or qubit, are currently under consideration~\cite{Ladd:2010}. Among these, superconducting qubits~\cite{Clarke:2008} occupy a leading position due to the potential for scalability~\cite{Kelly:2015, Gambetta:2017:AR} and robustness to dissipative phenomena~\cite{Barends:2013}. In fact, superconducting qubits can be fabricated on a chip using standard lithography techniques and can be operated with available microwave electronics. It has been shown that the error rates associated with the operation of these qubits can be as low as~\mbox{$\approx\!10^{-3}$}~\cite{Barends:2014, Jeffrey:2014}. While such rates are remarkable for a quantum-mechanical system, they are at least ten orders of magnitude higher than those necessary to run advanced quantum algorithms~\cite{Fowler:2012}.

It is believed that reaching the required error rates can be accomplished by means of quantum error correction~(QEC) algorithms~\cite{Gottesman:2010}, where quantum information is encoded in a large array of qubits. For example, a proof-of-concept implementation of the most forgiving QEC algorithm, the surface code~\cite{Fowler:2012}, can be realized with an ensemble of about~$100$ qubits. A similarly sized system may also make it possible to achieve quantum supremacy, i.e., to outperform the most advanced classical supercomputer, without resorting to QEC~\cite{Boixo:2017}.

The operation of a system with~$100$ or more superconducting qubits requires a complex classical infrastructure for qubit wiring, packaging, control, and measurement~\cite{Gambetta:2017:AR, Brecht:2016, Versluis:2017}. This infrastructure almost certainly leads to new sources of qubit error, such as correlated~\cite{Fowler:2014} and coherent errors~\cite{Orsucci:2016, Greenbaum:2016, Wood:2017}.

In this article, we study an extended version of the wiring and packaging technique introduced in~\cite{Bejanin:2016}: \textit{The large-scale quantum socket}. The qubit chip is housed in a microwave package that embeds a set of three-dimensional wires used to operate the qubits. The package behaves as a cavity with resonance modes~\cite{Lin:1994, Li:2000} that can interfere with qubit operation~\cite{Brecht:2016}. Quantum information can leak into these modes resulting in coherent errors~\cite{Wood:2017}.

We theoretically quantify coherent leakage errors by modeling the unwanted modes as a depolarizing channel. We propose two methods to mitigate coherent leakage and analyze the transition between coherent and incoherent errors by accounting for qubit damping (see section~\ref{sec:Methods}). We perform detailed simulations of the mitigation methods, showing that the three-dimensional wires can serve both to address qubits and isolate them from leakage into unwanted modes. We present preliminary experimental results demonstrating that a suitable placement of the wires inside a large-scale quantum socket can be used to create an electric shield against unwanted modes (see section~\ref{sec:Results}). Finally, we discuss a generalization of the mitigation methods (see section~\ref{sec:Discussion}) and outline the next steps for the implementation of a large-scale quantum socket with superconducting qubits (see section~\ref{sec:Conclusion}).

\section{Methods}
	\label{sec:Methods}

In this section, we summarize the main concepts of qubit coherent errors (see subsection~\ref{subsec:A:primer:to:qubit:coherent:errors}). We present a theoretical model to quantify the coherent leakage of superconducting qubits coupled to an unwanted cavity mode (see subsection~\ref{subsec:Theoretical:model:of:coherent:leakage:due:to:cavity:modes}). We introduce two methods to mitigate this type of coherent leakage: Half-wave fencing (see subsection~\ref{subsec:Half:wave:fencing}) and antinode pinning (see subsection~\ref{subsec:Antinode:pinning}). We describe the setup and settings of the electromagnetic field simulations performed to emulate an Xmon transmon qubit coupled to a cavity as well as the effects of three-dimensional wires on cavity modes (see subsection~\ref{subsec:Electromagnetic:field:simulations:setup:and:settings}). Finally, we present the experimental setup utilized for a proof-of-concept implementation of a large-scale quantum socket (see subsection~\ref{subsec:Experimental:setup}).

\subsection{A primer to qubit coherent errors}
	\label{subsec:A:primer:to:qubit:coherent:errors}

In a quantum computer, the Hilbert space of all qubits and any other internal auxiliary system required to operate them is defined as the \textit{computational subspace}; on the contrary, the space associated with any external system interacting with the qubits is called the \textit{leakage subspace}~\cite{Wood:2017}. In general, the time evolution of a qubit interacting with internal and external systems is described by the combination of a purely unitary generator~$\mathcal{H}$ and a purely dissipative generator~$\mathcal{D}$~\cite{Wood:2017, Rivas:2012}.

The generator~$\mathcal{H}$ accounts for the qubit Hamiltonian as well as the Hamiltonian of any wanted or unwanted internal system or any external system, or both. External systems always lead to unwanted dynamics and, thus, to qubit errors, whereas only unwanted internal systems generate qubit errors. The purely unitary nature of~$\mathcal{H}$ results in coherent dynamics, implying that all the errors associated with it are \textit{coherent errors}. In particular, errors due to external systems are called \textit{coherent leakage errors}. The generator~$\mathcal{D}$, instead, describes external environments acting as stochastic phenomena (e.g., Markovian noise). Therefore, the qubit errors associated with~$\mathcal{D}$ are defined as \textit{incoherent leakage errors}; these errors are typically due to qubit decoherence, i.e., relaxation and dephasing~\cite{Clarke:2008}. Note that two- or multi-qubit \textit{correlated errors} can also exist~\cite{Fowler:2014}. In this case, when an error occurs on one qubit it affects one or more different qubits in the quantum computer. Correlated errors can stem from either coherent or incoherent dynamics.

In superconducting qubit implementations, a typical example of a wanted internal system is a resonator acting as a quantum bus between pairs of qubits~\cite{Mariantoni:2011}. The states of the bus are populated during computations, although at the end of any computation only qubits states remain populated. A special class of wanted internal systems is represented by classical driving electromagnetic fields used to control and measure the qubit state. These systems result in unwanted dynamics when leading to stray fields that act on undesired qubits~\cite{Najafi-Yazdi:2017}. An example of an external system, instead, is a cavity mode due to the microwave package used to house a superconducting qubit chip. This mode can also generate correlated errors between two qubits that interact with it independently.

It has been shown that the implementation of a quantum socket for a~$10 \times 10$ array of Xmon transmon qubits necessitates a package with lateral dimensions of at least~$72 \times \SI{72}{\milli\meter\squared}$~\cite{Bejanin:2016}. A package of this size generally supports a set of unwanted cavity modes that can interact with the qubits, deteriorating the performance of the quantum computer significantly~\cite{Brecht:2016}. When considering the interaction between one qubit and one unwanted mode, the generator~$\mathcal{H} = - \rmi [ \widehat{H}_{\textrm{JC}} , \hat{\rho} ]$, where~$\widehat{H}_{\textrm{JC}}$ is the Jaynes-Cummings Hamiltonian~\cite{Haroche:2006}, $\hat{\rho}$ is the cavity-qubit density matrix, and $\rmi^2 = -1$; the generator~$\mathcal{D}$ accounts for qubit and cavity decoherence. The time evolution of~$( \mathcal{H} + \mathcal{D} )$ results in leakage errors. In this case, the interplay between the coherent and incoherent error regimes is largely dictated by the detuning between the qubit transition frequency~$f_{\textrm{q}}$ and the cavity mode resonance frequency~$f_{\textrm{c}}$, $\Delta = ( f_{\textrm{c}} - f_{\textrm{q}} )$, as well as by cavity and qubit damping rates.

\subsection{Theoretical model of coherent leakage due to cavity modes}
	\label{subsec:Theoretical:model:of:coherent:leakage:due:to:cavity:modes}

In order to study the dynamics of a cavity mode coupled to a qubit in presence of qubit decoherence, we resort to a generalized form of the Maxwell-Bloch equations~\cite{Arecchi:1965, Loudon:2000}. In the semiclassical approximation, the cavity mode Hamiltonian is represented by a classical monochromatic electromagnetic field with frequency~$f_{\textrm{c}}$. Setting the energy of the qubit ground~state~$\ket{\textrm{g}}$ to zero and naming~$\ket{\textrm{e}}$ the qubit energy excited state, the qubit Hamiltonian is~$\widehat{H}_{\textrm{q}} = h f_{\textrm{q}} \ket{\textrm{e}} \!\! \bra{\textrm{e}}$. The qubit density matrix reads
\begin{equation}
\hat{\rho}_{\textrm{q}} = \left[
	\begin{array}{cc}
		\rho_{\textrm{gg}} & \rho_{\textrm{ge}} \\
		\rho_{\textrm{eg}} & \rho_{\textrm{ee}}
	\end{array}
						  \right] ,
	\label{eq:hat:rho:q}
\end{equation}
where~$\rho_{\textrm{gg}} , \rho_{\textrm{ee}} \in \mathbb{R}_{\geq 0}$ are the qubit ground state and excited state populations, respectively, with \mbox{$\rho_{\textrm{gg}} + \rho_{\textrm{ee}} = 1$}, and $\rho_{\textrm{eg}} = \rho^{\ast}_{\textrm{ge}} \in \mathbb{C}$ are the qubit coherences. Finally, the cavity-qubit interaction Hamiltonian is assumed to be an electric-dipole Hamiltonian~$\widehat{H}_{\textrm{ed}}$ with coupling coefficient~$g = E_0 p_{\textrm{q}} / h$, where~\cite{Haroche:2006, Loudon:2000}
\begin{equation}
E_0 = \left( \frac{h f_{\textrm{c}}}{2 \epsilon_0 V} \right)^{1/2}
	\label{eq:E:0}
\end{equation}
is the zero-point electric field of a cavity with volume~$V = a \times b \times e$ and $p_{\textrm{q}}$ is the qubit effective electric dipole moment~\cite{Blais:2004} (here, $\epsilon_0$ is the electric permittivity of free space, $h$ the Planck constant, and $a , b, e$ the cavity dimensions). In our definition of~$g$, the qubit is assumed to be at an antinode of the electric field.

In any quantum computation, the system of a single qubit and an unwanted cavity mode is characterized by at most one energy excitation. In fact, a typical initial condition for the system dynamics is when the qubit is prepared in state~$\ket{\textrm{e}}$ and the cavity mode is in the vacuum state~$\ket{\textrm{0}}$. In this case, replacing the quantized cavity mode with a classical field of unit amplitude leads precisely to the same dynamics as a fully quantized Jaynes-Cummings model limited to the one excitation sector (neglecting any damping terms).

Qubit decoherence can be taken into account by including relaxation processes with rate~$\gamma_{\textrm{r}} = 1 / T_1$, where~$T_1$ is the qubit relaxation time, and dephasing processes with rate~$\gamma_{\textrm{d}} = 1 / T_2 = 1 / ( 2 T_1 ) + 1 / \tau_{\phi}$, where~$T_2$ and $\tau_{\phi}$ are the qubit dephasing and pure dephasing time, respectively~\cite{Clarke:2008}. Similarly, cavity decoherence can be included by defining a rate~$\kappa_{\textrm{c}}$, which accounts for both relaxation and dephasing (the latter being usually negligible compared to the former~\cite{Wang:2009}).

When~$\Delta \approx 0$ and $g > \max ( \gamma_{\textrm{r}} , \gamma_{\textrm{d}} , \kappa_{\textrm{c}} )$, a cavity-qubit system prepared in the one excitation sector leads to vacuum Rabi oscillations (i.e., semi-resonant swaps in the strong coupling regime)~\cite{Haroche:2006}. Note that, this is the most likely situation for a qubit resonantly coupled to a cavity mode of a microwave package (see subsection~\ref{subsec:Coherent:leakage:theory:and:mitigation:A:Realistic:example}).

When, instead, $\Delta \gg g$, the system is in the dispersive regime. For~$( \gamma_{\textrm{r}} , \gamma_{\textrm{d}} ) \gtrsim \kappa_{\textrm{c}}$~\footnote{Since the microwave package is typically made from a metal (e.g., aluminum) that is in the superconducting state at the operating temperature of the quantum computer~\cite{Bejanin:2016}, this condition is the most likely to be fulfilled in realistic implementations. Superconductivity results in a high quality factor and, thus, a low~$\kappa_{\textrm{c}}$.}, the qubit decoherence rate can be approximated by~\cite{Blais:2004}
\begin{equation}
\frac{1}{T_1} \simeq \frac{\gamma_{\textrm{r}} + \gamma_{\textrm{d}}}{2} + \left( \frac{g}{\Delta} \right)^2 \, \kappa_{\textrm{c}} \, ,
	\label{eq:1:T:1}
\end{equation}
where the last term on the right-hand side is the Purcell effect. Choosing, for example, $\Delta = 10 g$, (\ref{eq:1:T:1}) results in~$1 / T_1 \simeq ( \gamma_{\textrm{r}} + \gamma_{\textrm{d}} ) / 2$. In this case, the Purcell effect is negligible compared to the bare qubit decoherence rates. This is the scenario we study in the remainder of this section. For~$( \gamma_{\textrm{r}} , \gamma_{\textrm{d}} ) \ll \kappa_{\textrm{c}}$, e.g., $\kappa_{\textrm{c}} = 100 \, ( \gamma_{\textrm{r}} + \gamma_{\textrm{d}} ) / 2$, and assuming the same detuning, (\ref{eq:1:T:1}) results in~$1 / T_1 \simeq ( \gamma_{\textrm{r}} + \gamma_{\textrm{d}} )$. In this case, the Purcell effect begins to deteriorate the bare qubit decoherence rates. We briefly consider this scenario in subsection~\ref{sec:Discussion}.

After a rotating-wave approximation, the generalized Maxwell-Bloch equations can be written in the interaction picture by defining~$\tilde{\rho}_{\textrm{ge}} = \tilde{\rho}^{\ast}_{\textrm{eg}} = \exp ( \rmi \Delta t ) \rho_{\textrm{ge}}$, $\tilde{\rho}_{\textrm{gg}} = \rho_{\textrm{gg}}$, and $\tilde{\rho}_{\textrm{ee}} = \rho_{\textrm{ee}}$, where~$t \in \mathbb{R}_{\geq 0}$ is time. At absolute zero temperature, the resulting equations read
\begin{subnumcases}{}
\! \frac{\rmd}{\rmd t} \tilde{\rho}_{\textrm{ee}} \,\, = -\frac{1}{2} \, \rmi ( 2 \pi g ) \left( \tilde{\rho}_{\textrm{ge}} - \tilde{\rho}_{\textrm{eg}} \right) - \gamma_{\textrm{r}} \, \tilde{\rho}_{\textrm{ee}} \hspace{3.0mm} \label{eq:d:tilde:rho:ee:dt} \\ [1.5mm]
\! \frac{\rmd}{\rmd t} \tilde{\rho}_{\textrm{ge}} \,\, = \frac{1}{2} \, \rmi ( 2 \pi g ) \left( \tilde{\rho}_{\textrm{gg}} - \tilde{\rho}_{\textrm{ee}} \right) - \left[ \, \rmi ( 2 \pi \Delta ) + \gamma_{\textrm{d}} \right] \tilde{\rho}_{\textrm{ge}} \, . \hspace{3.0mm} \nonumber \\ [-1.0mm] \label{eq:d:tilde:rho:ge:dt}
\end{subnumcases}
Note that the differential equations for~$\tilde{\rho}_{\textrm{gg}}$ and $\tilde{\rho}_{\textrm{eg}}$ can readily be obtained from~(\ref{eq:d:tilde:rho:ee:dt}) and (\ref{eq:d:tilde:rho:ge:dt}), respectively, using the properties of the qubit density matrix elements. Hence, the generalized Maxwell-Bloch equations constitute a system of four first-order ordinary differential equations.

The probability that the qubit is in state~$\ket{\textrm{e}}$ is given by~$P_{\textrm{e}} = \bra{\textrm{e}} \hat{\rho}_{\textrm{q}} \ket{\textrm{e}} = \rho_{\textrm{ee}}$. By solving the generalized Maxwell-Bloch equations for the initial conditions~$\tilde{\rho}_{\textrm{ee}} ( t \! = \! 0 ) = 1$ and $\tilde{\rho}_{\textrm{ge}} ( t \! = \! 0 ) = 0$, we find the time evolution of~$\rho_{\textrm{ee}}$, i.e., $P_{\textrm{e}} ( t )$.

In order to estimate the depolarizing probability~$p$ due to a leakage error of a qubit into an unwanted cavity mode, we compute the time average of~$P_{\textrm{e}} ( t )$ over a time period~$T$. Thus, the depolarizing error probability is
\vspace{2.5mm}
\begin{equation}
p = 1 - \frac{1}{T} \, \int_{0}^{T} \rmd t \, P_{\textrm{e}} ( t ) \, .
	\label{eq:p}
\end{equation}
We note that, in the QEC jargon, the depolarizing error probability~$p$ as calculated in~(\ref{eq:p}) effectively represents a bit-flip error~\cite{Fowler:2012, Gottesman:2010}.

For example, when the qubit and cavity mode are on resonance, $\Delta = 0$, and assuming no damping, $\gamma_{\textrm{r}} = \gamma_{\textrm{d}} = 0 ( = \kappa_{\textrm{c}} )$, the solution to the Maxwell-Bloch equations can be expressed in analytical form~\cite{Loudon:2000}. In this case,
\begin{equation}
p_{\textrm{max}} = 1 - \lim_{T \rightarrow \infty} \frac{1}{T} \, \int_{0}^{T} \rmd t \, \sin^2 \left( \frac{g}{2} \, t \right) = 0.5 \, ,
	\label{eq:p:max}
\end{equation}
which is the theoretically highest attainable leakage error rate for an unwanted cavity-qubit system. In this scenario, the resulting error is a completely coherent leakage error.

\subsection{Half-wave fencing}
	\label{subsec:Half:wave:fencing}

\begin{figure*}[t!]
	\centering
	\includegraphics[]{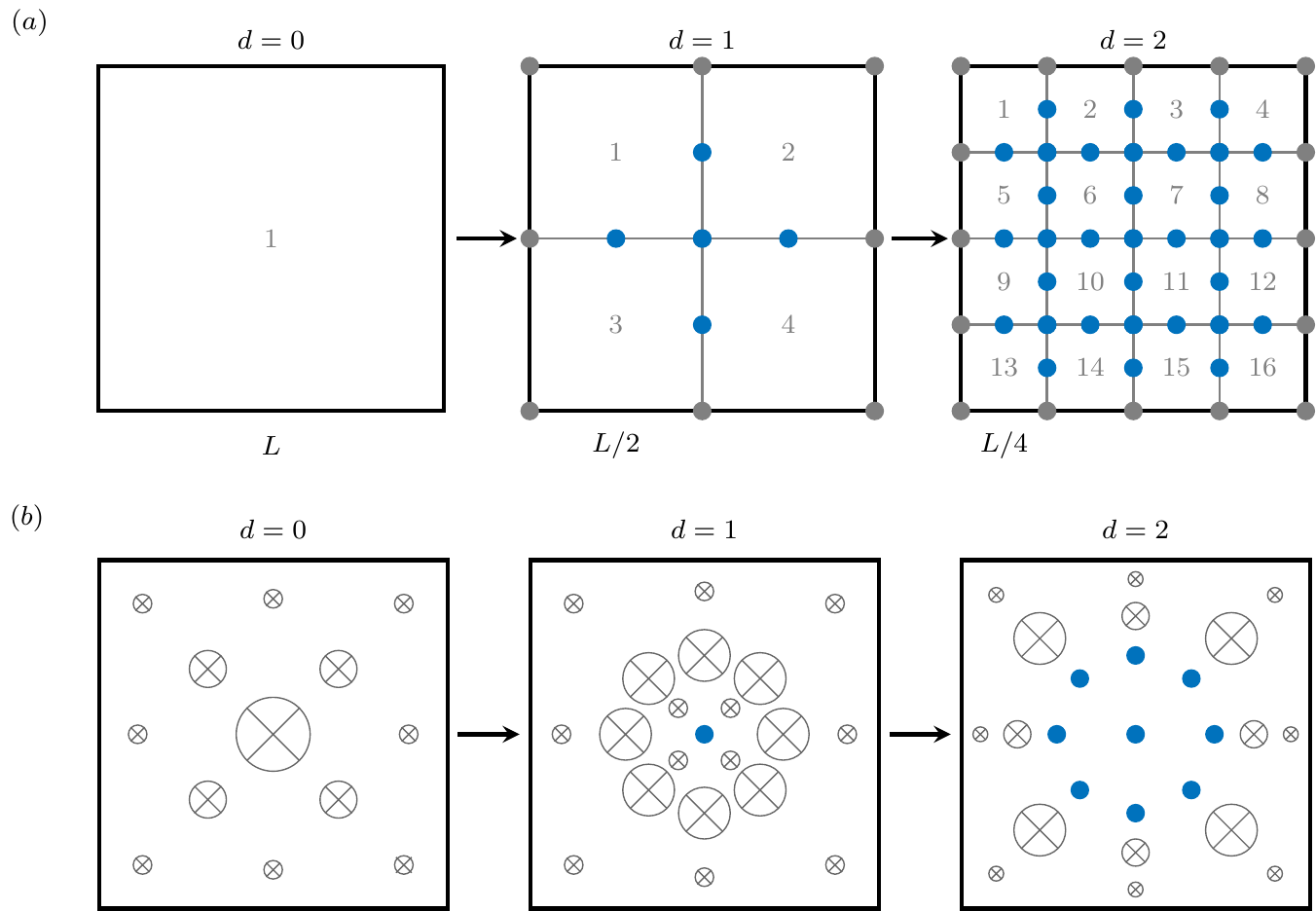}
\caption{First three iterations of the two proposed frequency shifting methods. The rectangular cavity of side~$L$ is viewed along the $y$-axis (i.e., top view). Three-dimensional wires are indicated by full blue circles. ($a$) Half-wave fencing. Numbers indicate each of the~$n_{\textrm{c}}$ cells for the various iterations. Gray lines represent the geometric boundary of a cell. Full gray circles represent the virtual location of wires in correspondence of actual cavity walls. ($b$) Antinode pinning. Crossed circles indicate the electric field distribution, with cross-circle size relating to the field strength (not to scale). For~$d = 0$, the~$\textrm{TE}_{1 0 1}$ mode is shown; for~$d = 1$, one wire pins the antinode of~$\textrm{TE}_{1 0 1}$, resulting in a toroidal field distribution; for~$d = 2$, $8$~additional wires partially pin the circular antinode. Note that, in this case, the field inside the~$8$-wire ``circle'' is very weak compared to the rest of the field and, thus, is not shown.}
	\label{figure1a_1b_mcconkey}
\end{figure*}

The inner space of a microwave package can be modeled by a rectangular cavity with dimensions~$a = e = L$ and $b = H$~\cite{Collin:2001}, where~$L$ and $H$ are the side length and height of a cavity with square cross section. The cavity supports both transverse electric~(TE) and transverse magnetic~(TM) modes with resonance frequencies
\begin{equation}
f_{n m \ell} = \frac{c}{2} \left( \frac{\ell^2 + n^2}{L^2} + \frac{m^2}{H^2} \right)^{1/2} ,
	\label{eq:f:nml}
\end{equation}
where $c$ is the velocity of light in vacuum and \mbox{$n , m , \ell \in \mathbb{N}$} refer to the number of half wavelengths spanned by the electromagnetic field in the $x$-, $y$-, and $z$-axis~\cite{Collin:2001}.

When~$L > H$, the mode with the lowest resonance frequency (or dominant mode) is the~$\textrm{TE}_{n m \ell} = \textrm{TE}_{1 0 1}$ mode. The corresponding resonance frequency is~$f_{1 0 1} = c / ( \sqrt{2} L )$. Under these conditions, the cavity can be represented by a two-dimensional square membrane with side~$L$, as shown in figure~\ref{figure1a_1b_mcconkey}~($a$). The~$L^2$~square can be iteratively divided into smaller squares with dimensions~$\{ ( L / 2 )^2 , ( L / 4 )^2 , \ldots , [ L / ( 2^d ) ]^2 \}$, where~$d \in \mathbb{N}$ is the number of iterations [see figure~\ref{figure1a_1b_mcconkey}~($a$)]. The total number of squares after $d$~iterations is thus~$n_{\textrm{c}} = 2^{2 d}$.

The physical implementation of this method is realized by dividing the cavity in a number~$n_{\textrm{c}}$ of smaller cavities, or \textit{cells}, which are encapsulated by perfectly conducting grounded walls; each cell behaves as a Faraday cage. The walls can be replaced by a large set of three-dimensional wires, resulting in a Faraday fence. Figure~\ref{figure1a_1b_mcconkey}~($a$) displays the fences for~$d = 0 , 1 , 2$. The iteration~$d = 1$, for example, comprises one vertical and one horizontal fence. In the simplest case, each of these two fences can be approximated by three equally separated wires, one of which is in the center of the $L^2$~cell (with the center wire being used only once). This method, which we name \textit{half-wave fencing}, makes it possible to shield pairs of adjacent cells. After~$d$ iterations, the dimensionless resonance frequency of the dominant mode for each of the $n_{\textrm{c}}$~cells and the total number of wires are given by
\begin{subnumcases}{}
\! \tilde{f}_{\textrm{c}} \,\, = \frac{f_{\textrm{c}}}{f_{1 0 1}} \, 2^d \hspace{13.0mm} \label{eq:tilde:f:c:d} \\ [1.5mm]
\! N = ( 2^d + 1 )^2 + 2 \! \times \! 2^d \, ( 2^d + 1 ) - 2 \! \times \! 4 \! \times \! 2^d , \hspace{13.0mm} \label{eq:N:d}
\end{subnumcases}
respectively, where~$N \in \mathbb{N}$ is obtained from a simple counting argument. As expected, $\tilde{f}_{\textrm{c}}$ increases with~$d$ (the smaller the cell, the higher the resonance frequency). Note that the last term in~(\ref{eq:N:d}) is required not to count the unnecessary wires in correspondence to the four edges of the $L^2$~cell. By substituting~(\ref{eq:tilde:f:c:d}) into (\ref{eq:N:d}) we obtain a quadratic equation in the variable~$\tilde{f}_{\textrm{c}}$, with parameter~$N$. By selecting the positive solution of this equation, we find an analytic expression for~$\tilde{f}_{\textrm{c}}$ as a function of~$N$, which reads
\vspace{2.5mm}
\begin{equation}
\tilde{f}_{\textrm{c}} = \frac{1}{3} \left( 2 + \sqrt{1 + 3 N} \right) .
	\label{eq:tilde:f:c:N}
\end{equation}
\vspace{2.5mm}
This equation makes it possible to find~$\tilde{f}_{\textrm{c}}$ for any integer or even real value of~$N$, resulting in an extension of the half-wave fencing method. In real implementations, however, the method performs optimally only for a number of wires given by~(\ref{eq:N:d}), which corresponds to an even number of fully fenced cells of equal size.

It is worth noting that fully fenced cells can be formed for values of~$N$ other than those obtained with~(\ref{eq:N:d}), depending on the exact procedure followed to insert each new wire (at least for~$N > 2$). It is simpler to consider a number of wires given by exactly~(\ref{eq:N:d}) when describing the interaction between a qubit and a cell to determine coherent leakage errors. In this case, the volume of one cell is~$V = a^2 H$, where~$a = L / 2^d$ and $H$ is maintained constant. The corresponding zero-point electric field for the lowest resonant mode of the cell is obtained from~(\ref{eq:E:0}) and reads
\begin{equation}
E_{\textrm{c} 0} = \frac{1}{c} \left( \frac{h f^3_{\textrm{c}}}{\epsilon_0 H} \right)^{1/2} \, .
	\label{eq:E:c:0}
\end{equation}
Note that this equation is valid up to a certain frequency cutoff dictated by the smallest cell size~\footnote{That is, $f_{\textrm{c}} \nrightarrow + \infty$.}.

\subsection{Antinode pinning}
	\label{subsec:Antinode:pinning}

An alternative method to increase the resonance frequency of the modes in a rectangular cavity can be realized by introducing new boundary conditions at the antinodes of the cavity electric field~$\vec{E}$, i.e., where~$\| \vec{E} \|$ is maximum. The boundary conditions can be physically implemented by means of three-dimensional wires, the conductive nature of which forces~$\| \vec{E} \| = 0$. Figure~\ref{figure1a_1b_mcconkey}~($b$) displays the first three iterations of the algorithm used to perform this method, which we name \textit{antinode pinning}.

The spatial distribution of the electric field associated with the~$\textrm{TE}_{1 0 1}$ mode is characterized by a sinusoidal shape; the field antinode is located at the center of the cavity. When introducing a wire at this location, iteration~$d = 1$, the dominant mode (and all other modes) increases in frequency and the field distribution no longer resembles that of a~$\textrm{TE}_{n m \ell}$ mode. In this case, $\| \vec{E} \|$ has a toroidal structure with a continuous antinode distribution of circular shape.

The three-dimensional wires, however, can only be placed at semi-discrete locations. Thus, when performing iteration~$d = 2$, it is only possible to partially pin the entire antinode distribution by means of a finite number of wires. The performance of the method improves with the number of wires, until reaching the limit where wires start touching each other.

In subsequent iterations, the specific wire placement for continuous antinode distributions significantly impacts the overall performance of the antinode pinning method, as discussed in subsection~\ref{subsec:Electromagnetic:field:simulations}. In addition, the cavity modes depart from a simple geometric structure that can be described with analytical functions. Under these conditions, the wire placement and corresponding electromagnetic field distribution must be determined through numerical simulations. These can be simplified by solving the two-dimensional wave equation, i.e., ignoring the \mbox{$y$-axis}, assuming the wires to be additional boundary conditions. It is possible to perform the antinode pinning method efficiently by executing the algorithm of figure~\ref{figure1a_1b_mcconkey}~($b$) automatically until a desired frequency or maximum number of wires is reached.

\subsection{Electromagnetic-field simulations setup and settings}
	\label{subsec:Electromagnetic:field:simulations:setup:and:settings}

In general, the half-wave fencing and antinode pinning methods are \textit{frequency shifting} methods. In order to study in detail the effects of these methods on unwanted cavity modes, we resort to numerical simulations of the electromagnetic field using the high-frequency three-dimensional
full-wave electromagnetic-field simulation
software (also known as HFSS) by Ansys, Inc.~\footnote{See http://www.ansys.com/products/electronics/ansys-hfss for details on~HFSS.}.

The typical model used in our simulations is an ideal cavity with dimensions~$L = \SI{72}{\milli\meter}$ and $H = \SI{3}{\milli\meter}$, which is simulated by means of the HFSS eigenmode solution type. When considering a large number of three-dimensional wires inside the cavity, the electric field distribution is initially unknown. Thus, we do not make use of any mesh operation at the beginning of the simulation, instead allowing the adaptive meshing procedure to determine the optimal meshing layout. A maximum delta frequency pass of~\SI{0.01}{\percent} is used, with a minimum converged pass count of~$3$. The maximum number of passes is modified between models to avoid element counts larger than two million, as this would exceed the computing capabilities of our machine.

In the case of the antinode pinning method, we solve for the dominant eigenmode of a certain model that is then overlaid with the resulting~$\| \vec{E} \|$. In order to pin the antinode of this eigenmode, we assume a perfectly conducting three-dimensional wire with given diameter (see subsection~\ref{subsec:Electromagnetic:field:simulations}). We note that the actual outer conductor of the wires used in the quantum socket is made from brass~\cite{Bejanin:2016}. However, the thickness of this conductor is orders of magnitude larger than the skin depth and, thus, the perfect conductor idealization can be safely used. After pinning the antinode, we solve again for the dominant eigenmode and repeat the procedure by placing new wires at each antinode of the new eigenmode.

The half-wave fencing method is simpler to study than the antinode pinning method because it can be simulated without previous knowledge of the electric field distribution at each iteration. As shown in subsection~\ref{subsec:Electromagnetic:field:simulations}, the last iteration to be simulated for the half-wave fencing method is~$d = 3$ due to the computational limitations of our machine.

\subsection{Experimental setup}
	\label{subsec:Experimental:setup}

\begin{figure*}[t!]
	\centering
	\includegraphics[]{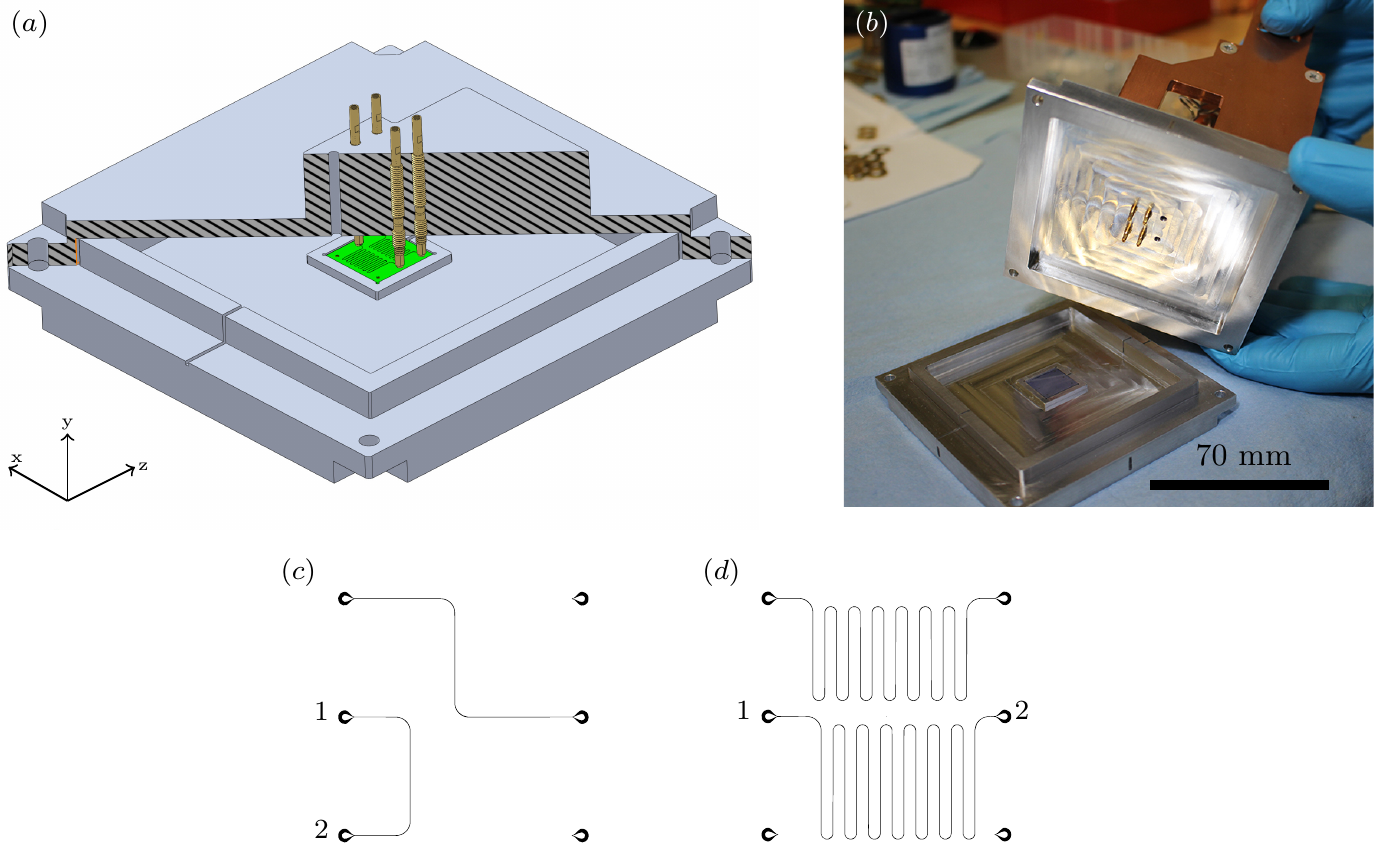}
\caption{The large-scale quantum socket. ($a$) Computer-aided design~(CAD) of the socket's lid and sample holder. A partial cutaway of the lid (hatching) reveals a set of four three-dimensional wires and a chip (green). ($b$) Image of the socket in~($a$) as implemented. ($c$)-($d$) Layout of the on-chip structures used in the transmission-coefficient measurements of subsection~\ref{subsec:Preliminary:experiments}, with input and output ports indicated by~$1$ and $2$, respectively. Both structures are CPW transmission lines with characteristic impedance~$Z_{\textrm{c}} \approx \SI{50}{\ohm}$, center conductor of width~$S \approx \SI{15}{\micro\meter}$, and gaps of width~$W \approx \SI{9}{\micro\meter}$. The lines have length~$\ell_{\tau} \approx \SI{12}{\milli\meter}$ [short line; ($c$)] and $\ell_{\tau} \approx \SI{85}{\milli\meter}$ [long line; ($d$)] from pad center to pad center. The pad design is described in~\cite{Bejanin:2016}.}
	\label{figure2a_2d_mcconkey}
\end{figure*}

The microwave package and package holder of the experimental setup used to implement a proof-of-concept large-scale quantum socket are shown in figure~\ref{figure2a_2d_mcconkey}~($a$) and ($b$). This setup allows us to assess the effects of the modes of a large cavity on a chip with superconducting structures. The internal cavity of the microwave package has dimensions~$L \approx \SI{72}{\milli\meter}$ and $H \approx \SI{6}{\milli\meter}$, making it significantly larger than our standard quantum sockets~\cite{Bejanin:2016} and most of the packages used in the superconducting qubit industry.

The chip under test resides in the sample holder, as shown in figure~\ref{figure2a_2d_mcconkey}~($a$) and ($b$). In order to reduce the presence of electromagnetic modes due to the chip's substrate, we restrict the substrate area to~$15 \times \SI{15}{\milli\meter\squared}$ with a thickness of~$\approx\!\SI{500}{\micro\meter}$. Any chip with bigger dimensions can generate modes interfering with the on-chip structures and possibly masking the cavity modes to be tested. We additionally include a small recess beneath the chip, thus forming the lower cavity described in~\cite{Bejanin:2016}. A small pillar made from polytetrafluoroethylene (PTFE) ``Teflon'' is inserted in the middle of this cavity for mechanical support of the chip. We estimate the dominant mode of the lower cavity to be at~$\approx\!\SI{13.5}{\giga\hertz}$ from simulations. The chip is encapsulated by a metallic protrusion that makes it possible to isolate the lower cavity from the inner cavity of the package.

The microwave package features four three-dimensional wires mounted in the middle of the lid and that mate with corresponding on-chip pads. The wire-pad pattern follows that in~\cite{Bejanin:2016}, allowing compatibility with chip designs that can be tested in our standard quantum sockets. In addition, the package holder is made to be compatible with the interconnects used in our dilution refrigerator setup.

The inner cavity of our standard quantum sockets can be emulated in the large-scale socket by means of a metallic filler. This reduces the dimensions of the inner cavity to~$L \approx \SI{15}{\milli\meter}$ and $H \approx \SI{6}{\milli\meter}$, allowing us to effectively compare large- and small-scale sockets with the same microwave package and chip.

The chips are fabricated by depositing a \SI{150}{\nano\meter} thick aluminum (Al) film followed by a \SI{150}{\nano\meter} thick indium (In) film on a \SI{500}{\micro\meter} thick silicon (Si) substrate. The films are patterned by means of optical lithography, forming the structures shown in figure~\ref{figure2a_2d_mcconkey}~($c$) and ($d$). We note that the In films are not necessary for the purposes of this project. However, thanks to the compatibility with our standard quantum sockets we have been able to use the same chips as a measurement reference in the study reported in~\cite{McRae:2017}, where In films are a key requirement. The patterned structures are coplanar-waveguide~(CPW) transmission lines of different lengths (see caption of figure \ref{figure2a_2d_mcconkey} for details).

Both the microwave package and filler are made by way of high-precision machining from Al 6061-T6. All measurements are performed at a temperature of approximately~\SI{10}{\milli\kelvin} in a dilution refrigerator. More details on sample fabrication and measurement setup are in~\cite{Bejanin:2016}.

\section{Results}
	\label{sec:Results}

In this section, we illustrate with a realistic example the theoretical model introduced in subsection~\ref{subsec:Theoretical:model:of:coherent:leakage:due:to:cavity:modes} and estimate the depolarizing error probability when using the method proposed in subsection~\ref{subsec:Half:wave:fencing} (see subsection~\ref{subsec:Coherent:leakage:theory:and:mitigation:A:Realistic:example}). We present electromagnetic field simulations of the frequency shifting methods of subsections~\ref{subsec:Half:wave:fencing} and \ref{subsec:Antinode:pinning} (see subsection~\ref{subsec:Electromagnetic:field:simulations}). Finally, we show a set of preliminary experimental results towards the implementation of a large-scale quantum socket (see subsection~\ref{subsec:Preliminary:experiments}).

\subsection{Coherent leakage theory and mitigation: A realistic example}
	\label{subsec:Coherent:leakage:theory:and:mitigation:A:Realistic:example}

\begin{figure*}[t!]
	\centering
	\includegraphics[]{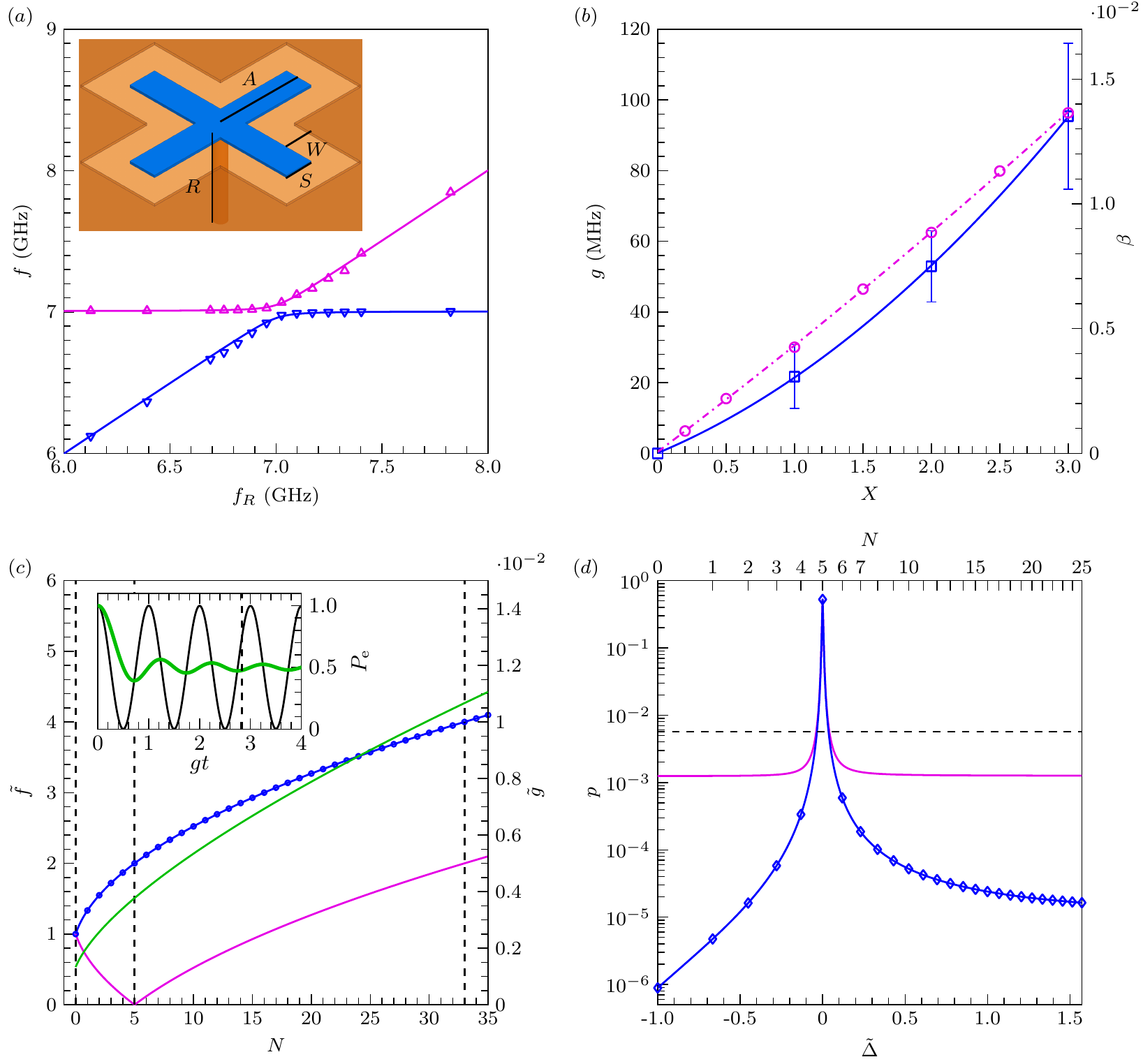}
\caption{Application of the coherent leakage theory and mitigation (see subsection~\ref{subsec:Coherent:leakage:theory:and:mitigation:A:Realistic:example}). ($a$) Simulation results of the first two eigenmodes ($f$ vs.~$f_R$) of a cross-cavity system for~$X = 3.0$, $\xi = \SI{100}{\micro\meter}$, and $\zeta = \SI{1000}{\micro\meter}$. The cavity dimensions are~$L = \SI{30.26}{\milli\meter}$ and $H = \SI{3}{\milli\meter}$, resulting in~$f_{\textrm{c}} \approx \SI{7}{\giga\hertz}$. The downward open blue triangles and the upward open magenta triangles correspond to~$\ket{- , 0}$ and $\ket{+ , 0}$, respectively. The solid lines are fitting curves obtained from~(\ref{eq:bar:f:mp:0}). Inset: CPW cross described in the main text for~$t_{\textrm{s}} = \SI{500}{\micro\meter}$ and $\epsilon_{\textrm{r}} = 11.45$ (Si at~$\approx \SI{4}{\kelvin}$). ($b$) Left $y$-axis: Values of $g$ obtained from the fitting procedure in~($a$) (open blue squares), with error bars indicating the~\SI{95}{\percent} confidence intervals of the fitting. The solid blue line is a quadratic fit. Right $y$-axis: $\beta$ vs.~$X$ (open magenta circles) with quadratic fit (dashed magenta line). ($c$) Left $y$-axis: Scaled frequency~$\tilde{f} = f / f_{1 0 1}$ vs.~$N$. $\tilde{f}_{\textrm{c}}$: Solid blue line; $| \tilde{\Delta} |$: Solid magenta line. The open blue circles indicate~$\tilde{f}_{\textrm{c}}$ for integer values of~$N$. Right $y$-axis: $\tilde{g}$ vs.~$N$. We choose~$f_{101} = \SI{3}{\giga\hertz}$ and $f_{\textrm{q}} = 2 f_{101} = \SI{6}{\giga\hertz}$. The vertical dashed black lines are in correspondence of~$d = 0, 1$, and $2$, respectively. Inset: $P_{\textrm{e}}$ vs.~$t$ for~$g ( N \! = \! 5 ) = \SI{11}{\mega\hertz}$ and $\Delta = 0$. The solid green line shows~$( 1-p )$ for increasing values of the integration time~$t$. The vertical dashed black line indicates the integration time~$T$ used to obtain the results in~($d$). The numerical integration is performed by means of the trapezoidal rule for~$2000$ time points. ($d$) $p$ vs.~$\tilde{\Delta}$ in absence of qubit damping (solid blue line) and for~$T_1 = \SI{100}{\micro\second}$ and $T_2 = \SI{50}{\micro\second}$ (solid magenta line). The open blue diamonds correspond to integer values of~$N$ (top $x$-axis). The horizontal dashed black line indicates~$p_{\textrm{th}} = \SI{0.57}{\percent}$~\cite{Fowler:2012}.}
	\label{figure3a_3d_mcconkey}
\end{figure*}

In order to calculate the depolarizing error probability~$p$ given by~(\ref{eq:p}) in a realistic quantum computing setting, we study a typical Xmon transmon qubit~\cite{Barends:2013} located inside a three-dimensional cavity~\cite{Paik:2011}. We emulate the Xmon transmon qubit by means of the microwave structure depicted in the inset of figure~\ref{figure3a_3d_mcconkey}~(a).

The structure comprises a CPW cross on the \mbox{$xz$-plane}, with the center bottom attached to a micro-coaxial half-wave resonator of length~$R$ that extends outside the cavity, along the $y$-axis. The cross acts as an antenna that couples to the cavity modes with resonance frequencies given by~(\ref{eq:f:nml}), which are set by the geometry of the cavity. By continuously varying~$R$, it is possible to sweep the resonance frequency of the resonator, $f_R$, resulting in a tunable resonator coupled to a set of fixed cavity modes. The cross lies on the same plane as the metallic bottom wall of the cavity, with a dielectric substrate directly below. The center of the cross is positioned at the antinode of the electric field of the dominant mode. The dimensions of the CPW cross are the center conductor width~$S$, the gap width~$W$, and the arm length~$A$. The substrate is characterized by a thickness~$t_{\textrm{s}}$ and relative electric permittivity~$\epsilon_{\textrm{r}}$.

We perform electromagnetic field simulations of the cross-cavity coupled system by means of ANSYS HFSS. We sweep~$f_R$ through~$f_\textrm{c}$, obtaining the first two eigenmodes shown in figure~\ref{figure3a_3d_mcconkey}~($a$). An accurate simulation of this system is computationally intensive due to the high aspect ratio between the largest and smallest feature of the system ($\sim\!\!\SI{10}{\milli\meter} / \SI{10}{\micro\meter}$). This issue can be overcome by scaling up the dimensions of the cross, while maintaining the same cavity size. The dimensions of each simulated cross are determined from~$S = W = \xi X$ and $A = \zeta X$, where~$X$ is a scaling factor and $\xi$ and $\zeta$ are two reference dimensions (see caption of figure~\ref{figure3a_3d_mcconkey} for numerical values). We simulate the coupled system for progressively smaller cross dimensions until exceeding the computing capabilities of our machine. Values of~$g$ for even smaller cross sizes can be extrapolated following the trend established by the simulated systems.

The first two eigenmodes of the electric field for a particular value of~$X$ are shown in figure~\ref{figure3a_3d_mcconkey}~($a$). This diagram resembles the energy level anti-crossing of a coupled cavity-qubit system. Thus, it can be used to estimate~$g$ by fitting the simulated frequency eigenmodes to the frequencies associated with the first two energy dressed states of the Jaynes-Cummings Hamiltonian, $\ket{- , 0}$ and $\ket{+ , 0}$, respectively, subtracted by the frequency of the ground state energy~\cite{Haroche:2006},
\begin{eqnarray}
\bar{f}_{\mp , 0} & = \frac{E_{\mp , 0} - E_{\textrm{g} , 0}}{h} \nonumber \\
& = \left( f_{\textrm{c}} \mp \frac{1}{2} \sqrt{g^2 + \Delta^2} - \frac{1}{2} \Delta \right) \, .
	\label{eq:bar:f:mp:0}
\end{eqnarray}
The curve fitting results are shown in figure~\ref{figure3a_3d_mcconkey}~($a$), where the Jaynes-Cummings model overlays the simulated data.

\begin{figure*}[t!]
	\centering
	\includegraphics[]{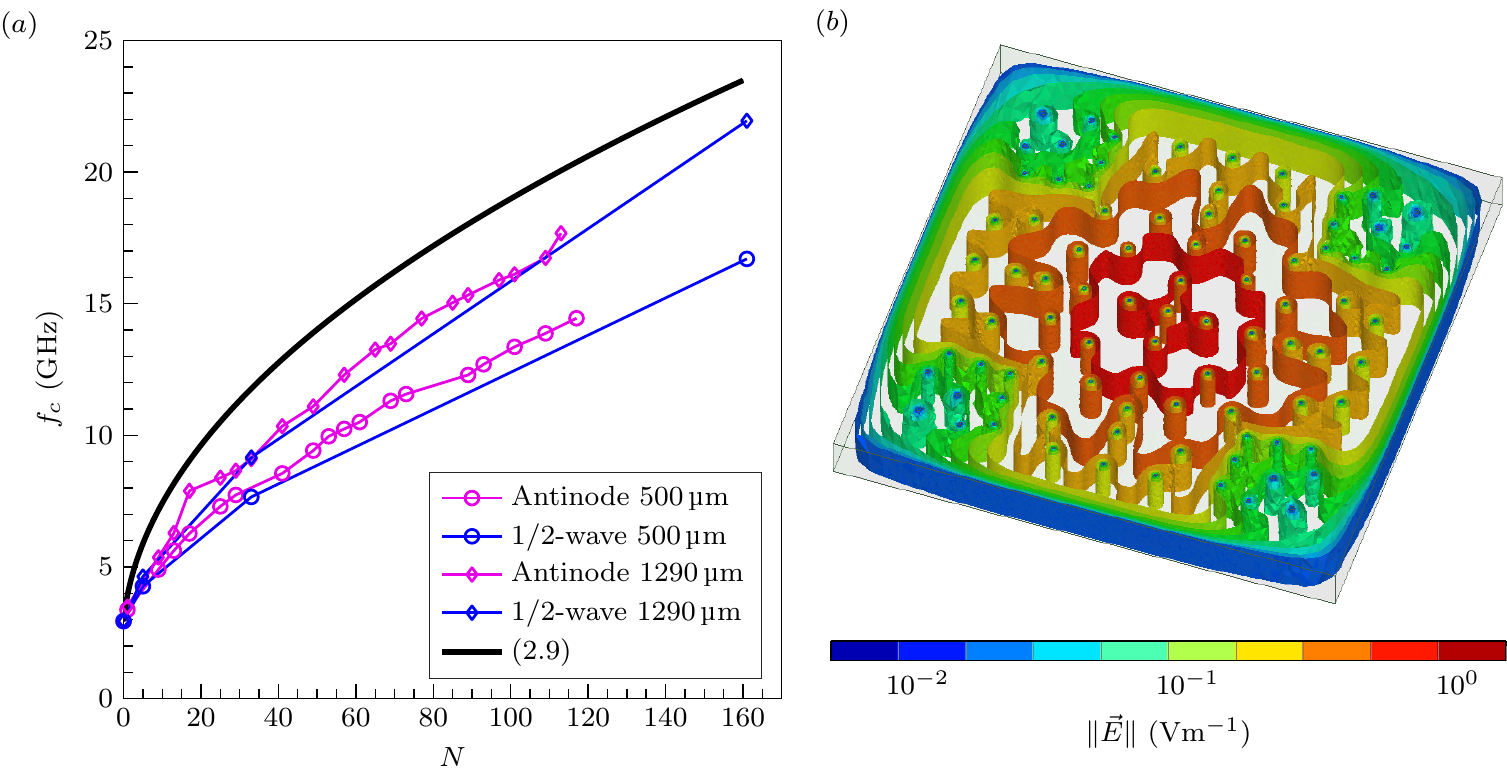}
\caption{Simulation results of the frequency shifting methods. ($a$) $f_{\textrm{c}}$ vs.~$N$ for the cavity dimensions given in subsection~\ref{subsec:Electromagnetic:field:simulations:setup:and:settings}, corresponding to~$f_{1 0 1} \approx \SI{2.944}{\giga\hertz}$. ``$1/2$-wave'': Half-wave fencing; ``Antinode'': Antinode pinning; wire dimensions indicated in the legend. The solid black line plots~(\ref{eq:tilde:f:c:N}), providing a pseudo-upper bound for all frequency shifting methods. ($b$) $\| \vec{ E } \|$ for the first resonant mode obtained with~$N = 89$ and wire diameter of~\SI{500}{\micro\meter} using the antinode pinning method; $f_\textrm{c} \approx \SI{12.295}{\giga\hertz}$.}
	\label{figure4a_4b_mcconkey}
\end{figure*}

We perform the simulation and curve fitting procedure for three different values of~$X$. The corresponding values of~$g$ are displayed in figure~\ref{figure3a_3d_mcconkey}~($b$) that also shows a quadratic fit of the data, allowing the estimation of~$g$ for other values of~$X$. The electric field sinusoidal distribution is taken into account in the simulations.

For the parameter space chosen in the simulations, our scaling argument and derived coupling values are consistent with those in~\cite{Brecht:2017}. In that study, it is shown that a change in transmon qubit geometry is equivalent to a change in the capacitance ratio~$\beta \approx C_{\textrm{g}} / ( C_{\textrm{g}} + C_{\textrm{p}} )$ for a constant cavity height. This ratio accounts for the capacitances~$C_{\textrm{g}}$ and $C_{\textrm{p}}$ between the CPW cross and the cavity top and bottom walls, respectively, and directly relates to a change in~$g$. We calculate~$\beta$ from~$C_{\textrm{g}}$ and $C_{\textrm{p}}$ obtained for the values of~$X$ used to find~$g$, as well as for additional intermediate values. The capacitances~$C_{\textrm{g}}$ and $C_{\textrm{p}}$ are simulated with ANSYS Q3D Extractor~\footnote{See http://www.ansys.com/Products/Electronics/ANSYS-Q3D-Extractor for details on~Q3D.}. The results are shown in figure~\ref{figure3a_3d_mcconkey}~($b$). The figure also displays a quadratic polynomial fit of the simulated data, allowing us to compare the relationship between~$\beta$ and $g$. In spite of a slight mismatch, the results are sufficiently accurate to provide an upper bound for the coupling coefficient of a typical Xmon transmon qubit.

Figure~\ref{figure3a_3d_mcconkey}~($c$) displays the scaled frequency~$\tilde{f}_{\textrm{c}}$ given by~(\ref{eq:tilde:f:c:N}), the absolute value of the scaled frequency detuning~$| \tilde{\Delta} | = | \Delta | / f_{101}$ for a given qubit transition frequency~$f_{\textrm{q}}$, as well as the scaled coupling coefficient~$\tilde{g} = g / f_{101}$ as a function of the number of three-dimensional wires~$N$. As expected, $\tilde{f}_{\textrm{c}}$ increases monotonically with the~$N$, demonstrating the working principle of the half-wave fencing method.

For the parameters chosen to obtain the graphs in figure~\ref{figure3a_3d_mcconkey}~($c$), iteration~$d = 1$ is realized for~$N = 5$, resulting in~$n_{\textrm{c}} = 4$. In this case, we can place a qubit in any of the four cells, where each cell is characterized by~$\tilde{f}_{\textrm{c}} = f_{\textrm{q}} / f_{101}$. This corresponds to the \textit{zero-detuning condition} for any of the four cell-qubit systems, $\tilde{\Delta} = 0$. For higher values of~$N$, $| \tilde{\Delta} |$ increases monotonically (similarly to~$\tilde{f}_{\textrm{c}}$) and a zero-detuning condition can never be reached again. This is due to the fact that we are considering the lowest resonant mode of each cell.

The~$N = 0$ value of~$g$ is extrapolated from the simulated values displayed in figure~\ref{figure3a_3d_mcconkey}~($b$) for a CPW cross with~$X = 0.2$, $g ( N \!\!\! = \!\!\! 0 ) \!\!\! \approx \!\!\! \SI{4}{\mega\hertz}$. This choice of~$g$ corresponds to a typical Xmon transmon qubit~\cite{Barends:2013}; it also represents the worst case scenario for coherent leakage errors, allowing us to find an upper bound for~$p$. As shown in figure~\ref{figure3a_3d_mcconkey}~($c$), $\tilde{g}$ increases monotonically with~$N$ due to the functional dependence of~(\ref{eq:E:c:0}) on~$f_{\textrm{c}}$ and, thus, on~$N$. This effect partially counteracts the benefits of the half-wave fencing method, as it increases the ratio~$g / \Delta$ for larger values of~$N$.

The inset of figure~\ref{figure3a_3d_mcconkey}~($c$) shows~$P_{\textrm{e}} ( t )$ for a qubit coupled to one of the four cells obtained when~\mbox{$N = 5$}. We position the qubit at the electric field antinode of that cell, resulting in a maximum cell-qubit coupling with coefficient~$g ( N \!\! = \!\! 5 )$. In this case, the cell-qubit dynamics leads to the highest coherent leakage error for that cell. In fact, the qubit information is swapped resonantly to the cell mode and vice versa. The choice of the integration time~$T$ depends on the specific quantum computing application. For example, when considering a QEC algorithm such as the surface code, a reasonable choice of~$T$ is the time length of one of the eight steps of the surface code cycle~\cite{Fowler:2012}. In this case, $p$ corresponds to the per-step error rate of the surface code. The typical length of a surface code cycle for superconducting qubit implementations is~$T_{\textrm{sc}} \approx \SI{2}{\micro\second}$~\cite{Kelly:2015}. We therefore choose~\mbox{$T = [ 8 / g ( N \! = \! 0 ) ] / 8 = \SI{250}{\nano\second}$}. When~$N = 5$, this corresponds to~$p \approx 0.5$ [see~(\ref{eq:p:max})].

Performing a similar calculation for values of~$N$ up to~$25$ results in the graphs of figure~\ref{figure3a_3d_mcconkey}~($d$), which displays~$p$ as a function of~$\tilde{\Delta}$. One of the graphs is obtained accounting for qubit decoherence, whereas the other is for a purely unitary evolution. The latter clearly demonstrates that the half-wave fencing method significantly mitigates coherent leakage errors. In particular, when~\mbox{$\Delta \approx 75 \, g ( N \! = \! 0 )$} the depolarizing probability~$p$ is dominated by incoherent errors due to qubit dissipation rather than coherent leakage. It is worth noting that in both cases, the condition $p < p_{\textrm{th}}$ is reached for~$\Delta \approx 38 \, g ( N \! = \! 0 )$, where~$p_{\textrm{th}}$ is the surface code per-step threshold error rate.

\subsection{Electromagnetic field simulations}
	\label{subsec:Electromagnetic:field:simulations}

\begin{figure*}[t!]
	\centering
	\includegraphics[]{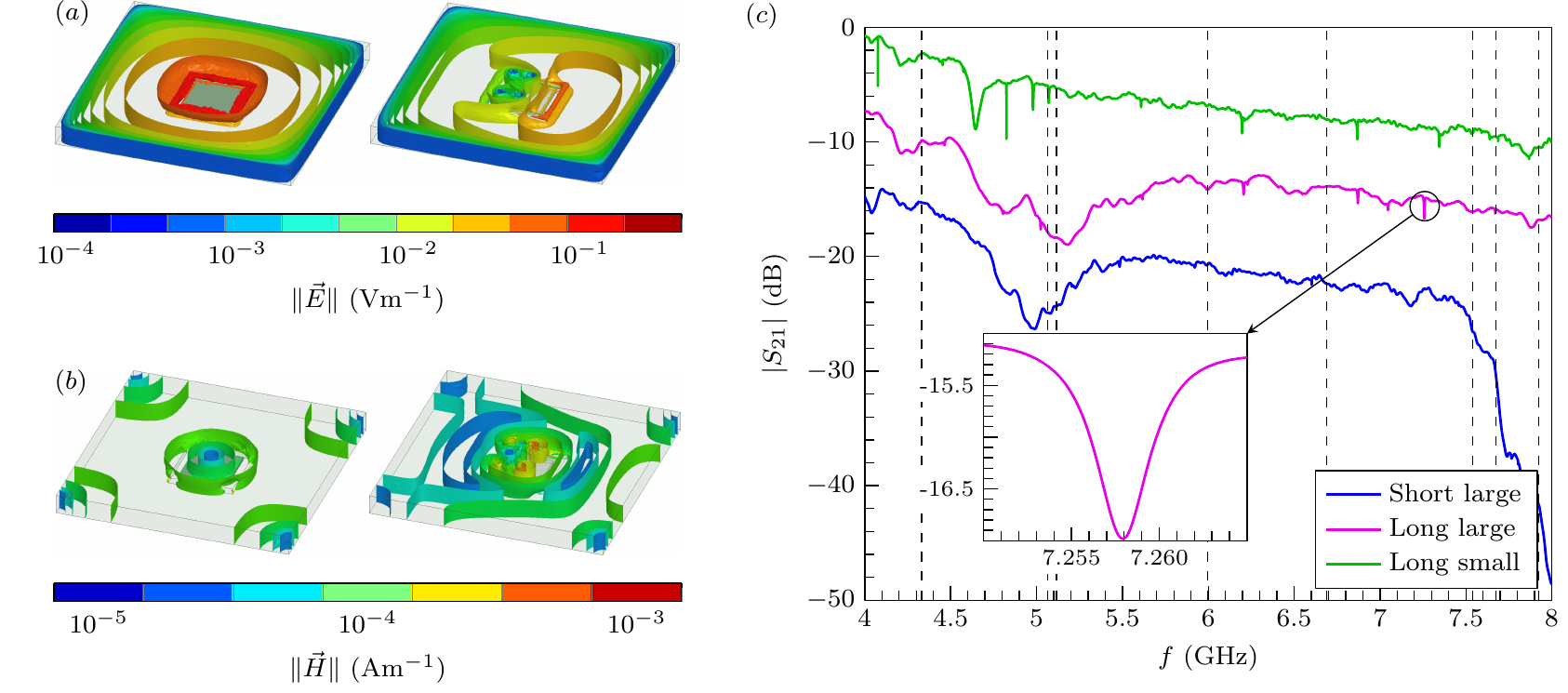}
\caption{Experimental results and comparison to simulations. ($a$)-($b$) Simulated~$\| \vec{ E } \|$ and $\| \vec{ H } \|$ for the microwave package shown in figure~\ref{figure2a_2d_mcconkey}~($a$) and ($b$) without~(left) and with~(right) four three-dimensional wires. Slight field bunching is noticeable around the chip fence. ($c$) $S_{21}$ vs.~$f$ for the structures drawn in figure~\ref{figure2a_2d_mcconkey}~($c$) and ($d$). All measurements are performed with a vector network analyzer with output power of~$\approx 0~\textrm{dB-milliwatts}$; the intermediate frequency bandwidth is typically~\SI{100}{\hertz}. ``Short large'', ``Long large'', and ``Long small'' indicate short and long transmission lines with~(large) and without~(small) filler (see subsection~\ref{subsec:Experimental:setup}). The vertical offset is due to measurement networks with different room-temperature amplifiers. The ``Short large'' rolloff above~$\approx \SI{7.5}{\giga\hertz}$ is due to a low-pass filter on the measurement network. The vertical dashed black lines indicate the frequencies reported in table~\ref{table:simulated:frequencies}. Inset: Zoom in on one of the deepest resonance features in the long transmission line measurement. Note that the feature is extremely small (less than~\SI{2}{\decibel}). From a semi lumped-element model, the small resonances are not related to the lines length~$\ell_{\tau}$ and may be due to slotline modes~\cite{Wenner:2011:a}.}
	\label{figure5a_5c_mcconkey}
\end{figure*}

The half-wave fencing method described by~(\ref{eq:tilde:f:c:N}) is based on the assumption of fully fenced cells. In reality, however, the cylindrical shape of the three-dimensional wires and their finite dimensions only allow for partial fencing. In order to more accurately estimate the effectiveness of this method as well as of antinode pinning, we must resort to numerical simulations of realistic scenarios. This allows us to tailor the design of large-scale quantum sockets to actual quantum computing applications, where it is necessary to reach a sufficiently high value of~$\Delta$ such that at least~$p < p_{\textrm{th}}/10$.

Figure~\ref{figure4a_4b_mcconkey} displays simulations of a large-scale quantum socket for three-dimensional wires of two different diameters. The wires with larger diameter correspond to those used in the microwave package shown in figure~\ref{figure2a_2d_mcconkey}, whereas the smaller wires are a future version planned to enable greater extensibility of the quantum socket.

The simulation results show a clear correlation between wire diameter and frequency shifting, where larger wires cause a larger shift for both methods. In the case of half-wave fencing, the correlation is due to the fact that larger wires better approximate an actual fence, where each cell is more isolated from its neighbors. In addition, wires with a larger diameter modify the ideal square shape of the cell, thus perturbing the corresponding electric field that, in turn, results in a higher~$f_\textrm{c}$. This effect becomes more prominent with higher values of~$d$ as the wire is proportionally bigger than each new cell size. This effect is also present in the case of antinode pinning, as bigger wires create larger new boundary conditions. As a consequence, the relative gap between~(\ref{eq:tilde:f:c:N}) and the simulation results reduces with~$N$.

We note that the antinode pinning method provides a slight advantage over the half-wave fencing method. However, the advantage is significant only when using wires with smaller dimensions. In fact, even an infinitesimally small wire would be sufficient to pin an antinode, provided the perfect conductor assumption remains fulfilled.

Figure~\ref{figure4a_4b_mcconkey}~($b$) displays an example of~$\| \vec{ E } \|$ for a cavity partially filled with three-dimensional wires. The electric field distribution clearly shows a situation where the optimal placement of the wires is made difficult by the complicated spatial distribution of the antinode.

\subsection{Preliminary experiments}
	\label{subsec:Preliminary:experiments}

\begin{table}[b!]
\caption{Simulated resonance frequency in gigahertz for the first eight modes of the microwave package shown in figure~\ref{figure2a_2d_mcconkey}~($a$) and ($b$). ``Without'': Simulation without the four three-dimensional wires. ``With'': Simulation with the four three-dimensional wires. ``Ratio'': Ratio between~$\| \vec{ E } \|$ near the center of the chip and at the antinodes. Lower values indicate weaker coupling between the on-chip structures and the cavity modes.}
	\label{table:simulated:frequencies}
\footnotesize
\begin{tabular*}{\columnwidth}{@{}l*{5}{@{\extracolsep{0pt plus
6pt}}l}}
\br
	& $f_1$ & $f_2$ & $f_3$ & $f_4$ \\
\mr
	Without & $2.77$ & $4.73$ & $4.73$ & $5.94$ \\
	With & $4.33$ & $5.06$ & $5.12$ & $6.00$ \\
	Ratio & $10^{-2}$ & $10^{-2}$ & $10^{-3}$ & $10^{-4}$ \\
\mr
	& $f_5$ & $f_6$ & $f_7$ & $f_8$ \\
\mr
	Without & $6.35$ & $6.66$ & $7.51$ & $7.51$ \\
	With & $6.69$ & $7.54$ & $7.67$ & $7.92$ \\
	Ratio & $10^{-1}$ & $5 \times 10^{-2}$ & $10^{-4}$ & $10^{-1}$ \\
\br
\end{tabular*}
\end{table}
\normalsize

The theory and simulations studied in this article have to be verified by means of experimental measurements on simple test structures. This testing stage is necessary to implement a large-scale quantum socket for actual quantum computing applications.

Figure~\ref{figure5a_5c_mcconkey}~($a$) and ($b$) displays the magnitude of the electric field~$\| \vec{ E } \|$ and magnetic field~$\| \vec{ H } \|$ for the large-scale quantum socket shown in figure~\ref{figure2a_2d_mcconkey}, both in the presence and absence of three-dimensional wires. It is worth noting that the introduction of the wires not only shifts the cavity mode frequencies upwards, but also reduces the electric field strength in proximity of the wires; conversely, the magnetic field generally increases in the same regions.

Table~\ref{table:simulated:frequencies} shows a comparison between the simulated resonance frequencies of the cavity modes with and without three-dimensional wires. In addition, it reports the relative magnitudes of~$\| \vec{ E } \|$ near the center of the chip for the first eight modes of the cavity in presence of wires.

Figure~\ref{figure5a_5c_mcconkey}~($c$) displays a set of preliminary experimental results, where the magnitude of the transmission coefficient~$| S_{21} |$ for the test structures in figure~\ref{figure2a_2d_mcconkey}~($c$) and ($d$) is plotted as a function of frequency. In our measurements, the cavity modes behave as if they were resonators capacitively or inductively coupled to the transmission line under test. Hence, they are expected to appear as dips in the transmission measurement.

In the case of the short transmission line no significant dips are visible across the entire frequency range. This is likely due to the fact that the line is located entirely within the four wires, where~$\| \vec{ E } \|$ is very small. In this case, the four wires effectively generate a cell of the half-wave fencing method. Even though~$\| \vec{ H } \|$ is slightly higher in this region, inductive coupling is typically much weaker than capacitive coupling and, thus, we expect no dips associated with~$\| \vec{ H } \|$.

The long (meandered) transmission line occupies a much larger area, including regions outside the wires where the electric field is much stronger. The transmission measurement shows a set of small dips, the center frequency of which, however, is far detuned from the expected values reported in table~\ref{table:simulated:frequencies}. In order to further confirm these dips are not due to cavity modes, we measure the same structure after introducing the filler described in subsection~\ref{subsec:Experimental:setup}. In this case, the dominant mode of the filled cavity is at~$\approx \SI{14.130}{\giga\hertz}$. The transmission measurement is very similar to the case without filler. It is therefore highly unlikely that these dips are due to cavity modes. This is per se an interesting result, as cavity modes have been identified to be a large source of interference in much smaller packages~\cite{Wenner:2011:a}.

\section{Discussion}
	\label{sec:Discussion}

The choice of the optimal frequency shifting method depends on a variety of factors: The qubits operation frequency; the dominant mode frequency of the package's bare cavity; the number of input and output lines required to control and measure the qubits; the size of the three-dimensional wires. In addition, any constraints on the qubit circuit layout can impact the wire placement.

For example, a surface code architecture comprising an array of~$10 \times 10$ qubits can be realized by means of a large-scale quantum socket with approximately~$250$ three-dimensional wires and a package's cavity with dimensions~$L = \SI{72}{\milli\meter}$ and $H = \SI{3}{\milli\meter}$, as the one used in the simulations of figure~\ref{figure4a_4b_mcconkey}. In this scenario, the required number of wires is sufficient to shift the dominant mode frequency far enough from the typical qubit operation frequency of~\SI{6}{\giga\hertz} such that~$p$ is at least two orders of magnitude below~$p_{\textrm{th}}$. This result is achieved with either frequency shifting method and for any realistically sized wire. For larger quantum sockets, it may be necessary to include ancillary wires solely devoted to frequency shifting (i.e., not used for qubit control and measurement).

Naturally, the half-wave fencing method is well suited to grid-type architectures such as that underlying the surface code. For a~$10 \times 10$ qubit array, the number of wires calculated from~(\ref{eq:N:d}) does not match the required~$250$ wires for any value of~$d$. However, the half-wave fencing method can be generalized to a method where the cavity side~$L$ is divided by $n > 1$. In this case, the quantity~$2^d$ in~(\ref{eq:tilde:f:c:d}) and (\ref{eq:N:d}) has to be substituted by~$n^d$, and $n_{\textrm{c}} = n^{2 d}$. Notably, the functional dependence of~$\tilde{f}_{\textrm{c}}$ on~$N$ is given by~(\ref{eq:tilde:f:c:N}) for any value of~$n$. Following this approach it is possible to wire up any~$n \times n$ qubit array, while simultaneously mitigating coherent leakage errors.

If the available number of wires is the limiting resource, antinode pinning is the ideal method of choice. In fact, it typically results in the greatest return on wire count, particularly when using small wires. Additionally, this is the most appropriate method when the constraints on the circuit layout are very restrictive. Suppose, for example, the user must initially place a set of wires at specific locations, ignoring any frequency shifting method. This scenario can be treated as the~$d = 0$ iteration of the antinode pinning method, thus making the method suitable for cavities of arbitrary shape.

It is also worth noting that the two frequency shifting methods are not mutually exclusive. Instead, they can be combined depending on the user requirements. For example, if the generalized half-wave fencing method only partially meets the wire requirements of an arbitrary two-dimensional array of qubits, the wiring can be completed by means of antinode pinning.

We note that the effect of the frequency shifting methods on higher cavity modes must also be taken into account. An example of this effect is shown in table~\ref{table:simulated:frequencies}, where the frequency shift of higher modes is less pronounced than for the dominant mode. This can result in a qubit being dispersively coupled to, e.g., $m$ modes close to each other in frequency. These modes act as multiple independent leakage channels. Assuming a depolarizing error probability~$p$ for each mode, the total error probability can be estimated to be~$m p$. According to the results in figure~\ref{figure4a_4b_mcconkey}~($d$), we can easily tolerate up to~$m \approx 100$, even though this will likely never happen in real applications.

In subsection~\ref{subsec:Theoretical:model:of:coherent:leakage:due:to:cavity:modes}, we only consider the strong coupling regime of a cavity-qubit system. Suppose, instead, that~$g < \max ( \gamma_{\textrm{r}} , \gamma_{\textrm{d}} , \kappa_{\textrm{c}} )$ (weak coupling regime). This regime suggests an alternative approach to manage unwanted cavity modes, where the microwave package is purposely made from low-quality materials, resulting in a high~$\kappa_{\textrm{c}}$. In the semi-resonant case, there are no coherent errors as the cavity-qubit dynamics is highly damped. However, in the dispersive regime care must be taken to insure that the Purcell effect does not dominate over the bare qubit decoherence rates. This can be achieved using the frequency shifting methods proposed here, following the qualitative recipe provided in~\cite{Abraham:2015:purcell}. Under these conditions, both the strong and weak coupling approaches lead to a similar result as that described by the magenta line in figure~\ref{figure3a_3d_mcconkey}~($d$).

An unwanted cavity mode can also mediate interactions between pairs of (or even multiple) qubits. As explained in subsection~\ref{subsec:A:primer:to:qubit:coherent:errors}, the corresponding dynamics can lead to correlated errors. The frequency shifting methods introduced here allow the separation of the qubits transition frequency from the cavity resonance frequency such that the qubit-cavity-qubit interaction gives rise only to virtual transitions. This result is similar to the dispersive two-qubit~$\sqrt{\rmi \textsc{SWAP}}$ gate introduced in~\cite{Blais:2004}. Assuming both qubits are coupled with the same coupling coefficient~$g$ to the cavity mode, the worst case scenario is when the qubits are in resonance with each other and detuned by~$\Delta$ from the mode. The effective qubit-qubit coupling strength is then~$g^2 / \Delta$ (dispersive coupling), which is strongly suppressed for values of~$\Delta$ larger than several times~$g$. In fact, such a dispersive coupling follows a similar dependence on~$\Delta$ as the dynamics leading to the plots in figure~\ref{figure3a_3d_mcconkey}~($c$) and ($d$).

Our frequency shifting methods are directly applicable to the problem of dielectric substrate modes, which has been addressed qualitatively in~\cite{Abraham:2014:b}. In this case, the resonance frequency of the substrate modes is given by~(\ref{eq:f:nml}) replacing~$c$ with~$c / \sqrt{\epsilon_{\textrm{r}}}$, where~$\epsilon_{\textrm{r}}$ is the relative electric permittivity of the substrate. The three-dimensional wires must be replaced by superconducting vias~\cite{Versluis:2017, Vahidpour:2017}, with all other methodological requirements remaining unchanged. The number of vias embedded in the substrate will need to be significantly higher than the number of wires in free space due to the lower frequency of the dominant mode. Notably, fabricating a large array of vias is a relatively simple process that can be made compatible with standard qubit fabrication techniques.

\section{Conclusion}
	\label{sec:Conclusion}

In this work, we study theoretically and with simulations a category of errors, coherent leakage errors, which becomes increasingly important with larger quantum computing architectures. We introduce a large-scale quantum socket based on three-dimensional wires that allows the operation of~$100$ or more superconducting qubits. We propose two methods, half-wave fencing and antinode pinning, that allow us to reduce the effect of unwanted cavity modes by means of the same wires used for qubit control and measurement. For example, the~$250$ wires required to operate a~$100$ qubit system make it possible to reach coherent leakage error rates of~$\approx 10^{-5}$, which are significantly lower than the rates due to incoherent errors.

There may be a maximum box size for which the frequency shifting methods will no longer efficiently reduce coherent leakage errors. However, this possible limitation can be overcome by adopting a modular architecture as the one proposed in~\cite{Brecht:2016}, where large-scale quantum sockets with hundreds of qubits are coupled together to form systems with thousands of qubits. It is worth noting that our frequency shifting methods can also be used to mitigate the effects of chip modes, another important source of errors in extensible superconducting qubit architectures.

The experimental results shown in this article indicate that the effect of box modes is strongly dependent on the chip design. In future projects, it is necessary to study large-scale quantum sockets using superconducting qubit chips. This will allow us to verify the effectiveness of the frequency shifting methods in a realistic quantum computing scenario.

\section*{Acknowledgements}

This research was undertaken thanks in part to funding from the Canada First Research Excellence Fund~(CFREF), the Discovery and Research Tools and Instruments Grant Programs of the Natural Sciences and Engineering Research Council of Canada~(NSERC), the Ministry of Research and Innovation~(MRI) of Ontario, and the Alfred P Sloan Foundation. We would like to acknowledge the Canadian Microelectronics Corporation~(CMC) Microsystems for the provision of products and services that facilitated this research, including CAD software and HFSS, as well as the Quantum NanoFab Facility at the University of Waterloo. MM acknowledges his fruitful discussions with Joseph Emerson and Austin G Fowler. We thank Evan A Peters for his comments on the manuscript.

\vspace{5.0mm}

\section*{References}

%\bibliography{largescalequantumsocket_mcconkey_bibliography}
%\bibliographystyle{iopart-num}
\providecommand{\newblock}{}

\end{document}